\documentstyle[floats,aps,aipbig,epsf]{revtex}
\begin{document}
\begin{flushright}
\Large
Fermilab-CONF-95/152-E \\
CDF/PUB/EXOTIC/PUBLIC/3192 \\
\today \\
\end{flushright}
\vspace*{0.1in}
\begin{center}
{\bf \LARGE Search for New Particles Decaying to Dijets,\\ $b\bar{b}$,
and $t\bar{t}$ at CDF\\}
\vspace*{0.2in}
{\Large CDF Collaboration\\ \vspace*{0.2in} Presented by \\Robert M. Harris\\}
\vspace*{0.1in}
{\Large \em Fermilab MS 318\\ Batavia, IL  60510\\}
\end{center}
\vspace*{-0.1in}
\large
\begin{abstract}
\large
We present three searches for new particles at CDF.  First, using 70 pb$^{-1}$
of data we search the dijet mass spectrum for resonances. There is an upward
fluctuation near 550 GeV/c$^2$ (2.6$\sigma$) with an angular distribution that
is adequately described by either QCD alone or QCD plus 5\% signal.  There
is insufficient evidence to claim a signal, but we set the most stringent
mass limits on the hadronic decays of axigluons, excited quarks, technirhos,
W$^{\prime}$, Z$^{\prime}$, and E6 diquarks.  Second, using 19 pb$^{-1}$ of
data we search the b-tagged dijet mass spectrum for $b\bar{b}$ resonances.
Again, an upward fluctuation near 600 GeV/c$^2$ (2 $\sigma$) is not significant
enough to claim a signal, so we set the first mass limits on topcolor bosons.
Finally, using 67 pb$^{-1}$ of data we search the top quark sample for
$t\bar{t}$ resonances like a topcolor $Z^{\prime}$.  Other than an
insignificant shoulder of 6 events on a
background of 2.4 in the mass region 475-550 GeV/c$^2$, there is no evidence
for new particle production.  Mass limits, currently in progress, should be
sensitive to a topcolor Z$^{\prime}$ near 600 GeV/c$^2$.  In all three searches
there is insufficient evidence to claim new particle production, yet there is
an exciting possibility that the upward fluctuations are the first signs
of new physics beyond the standard model.
\end{abstract}


%
\section{\large \bf Search for New Particles Decaying to Dijets}
\begin{figure}[t]
\epsfysize=7.5in
\epsffile[30 124 580 760]{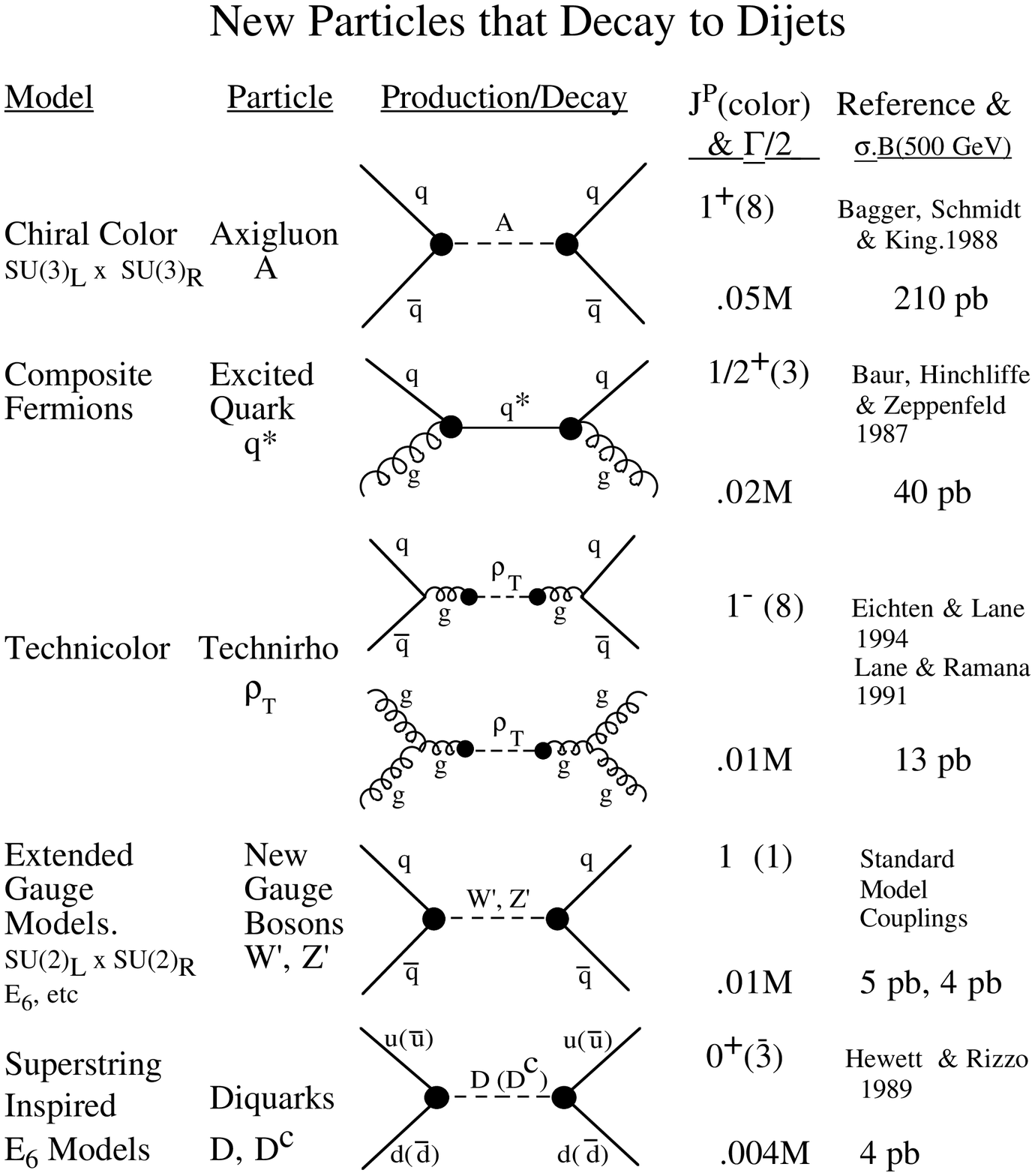}
\caption{ \large For each new particle that decays to dijets we list the model
name,
the particle name, the Feynman diagram, and a 2$\times$2 text grid containing
the quantum numbers, a reference, the half-width and the cross section at a
mass of 500 GeV/c$^2$.  The production and decay couplings for the first three
particles are strong, for new gauge bosons the coupling is weak, and
for $E6$ diquarks the coupling is electromagnetic.}
\label{fig_particles}
\end{figure}

As in our previous analysis of Run 1A data~\cite{ref_dijet}, we conduct
a general search for new particles with a narrow natural width that decay to
dijets. In addition, we search for the following particles summarized in
Fig.~\ref{fig_particles}:
axigluons~\cite{ref_axi} from
chiral QCD ($A \rightarrow q\bar{q}$),
excited states~\cite{ref_qstar} of composite quarks ($q^* \rightarrow qg$),
color octet
technirhos~\cite{ref_trho} ($\rho_T \rightarrow g \rightarrow q\bar{q},gg$),
new gauge bosons ($W^\prime$,$Z^\prime \rightarrow q\bar{q}$), and
scalar $E_6$ diquarks~\cite{ref_diquark} ($D\rightarrow \bar{u}\bar{d}$ and
$D^c \rightarrow ud$).

Using four triggers from run 1A and 1B, we combine dijet mass spectra
above a mass of 150 GeV/c$^2$, 241 GeV/c$^2$, 292 GeV/c$^2$, and 388 GeV/c$^2$
with integrated luminosities of
\clearpage
\begin{figure}[tbh]
\epsfysize=3.3in
\epsffile[30 144 540 680]{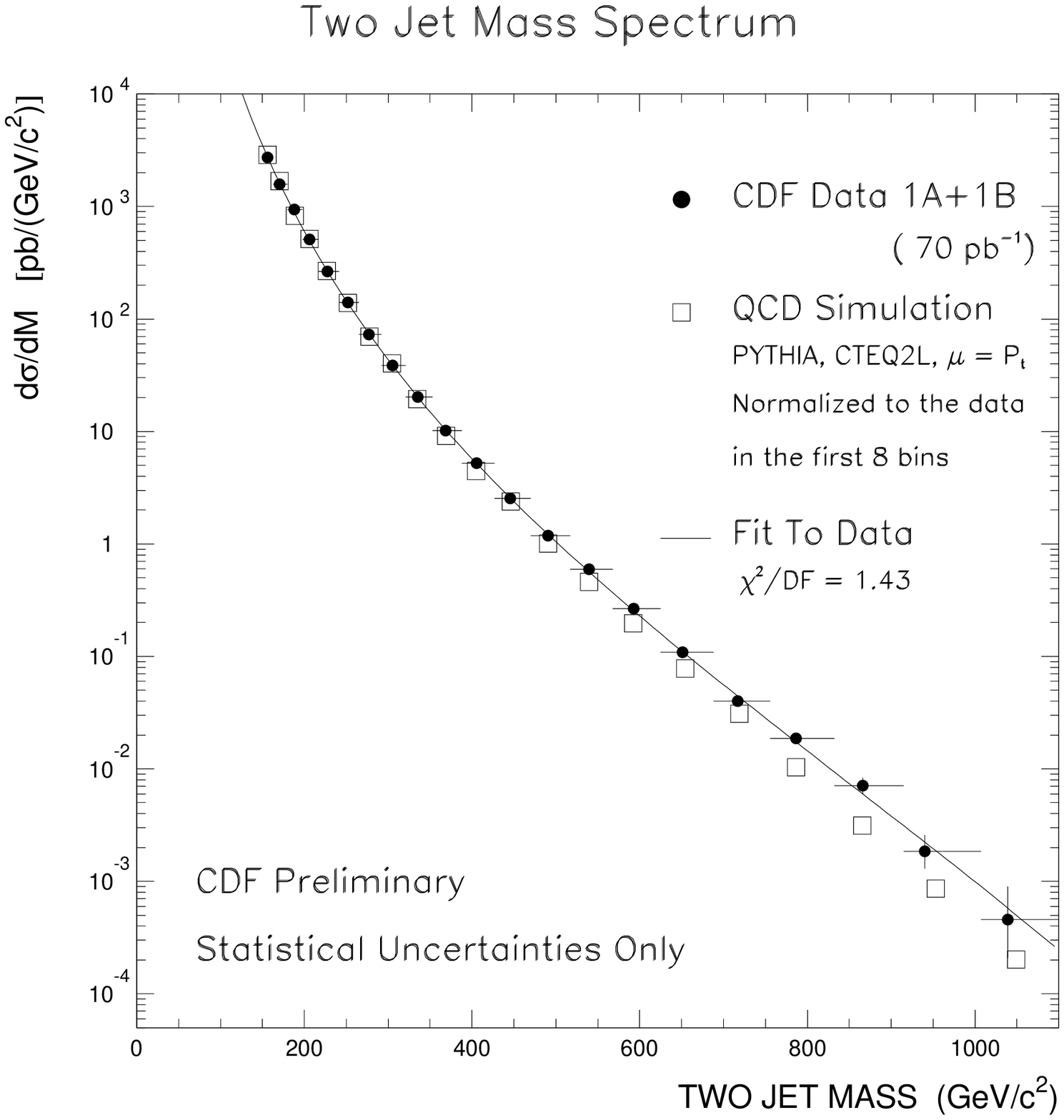}
\vspace*{-3.3in}
\hspace*{3.4in}
\epsfysize=3.3in
\epsffile[36 144 540 680]{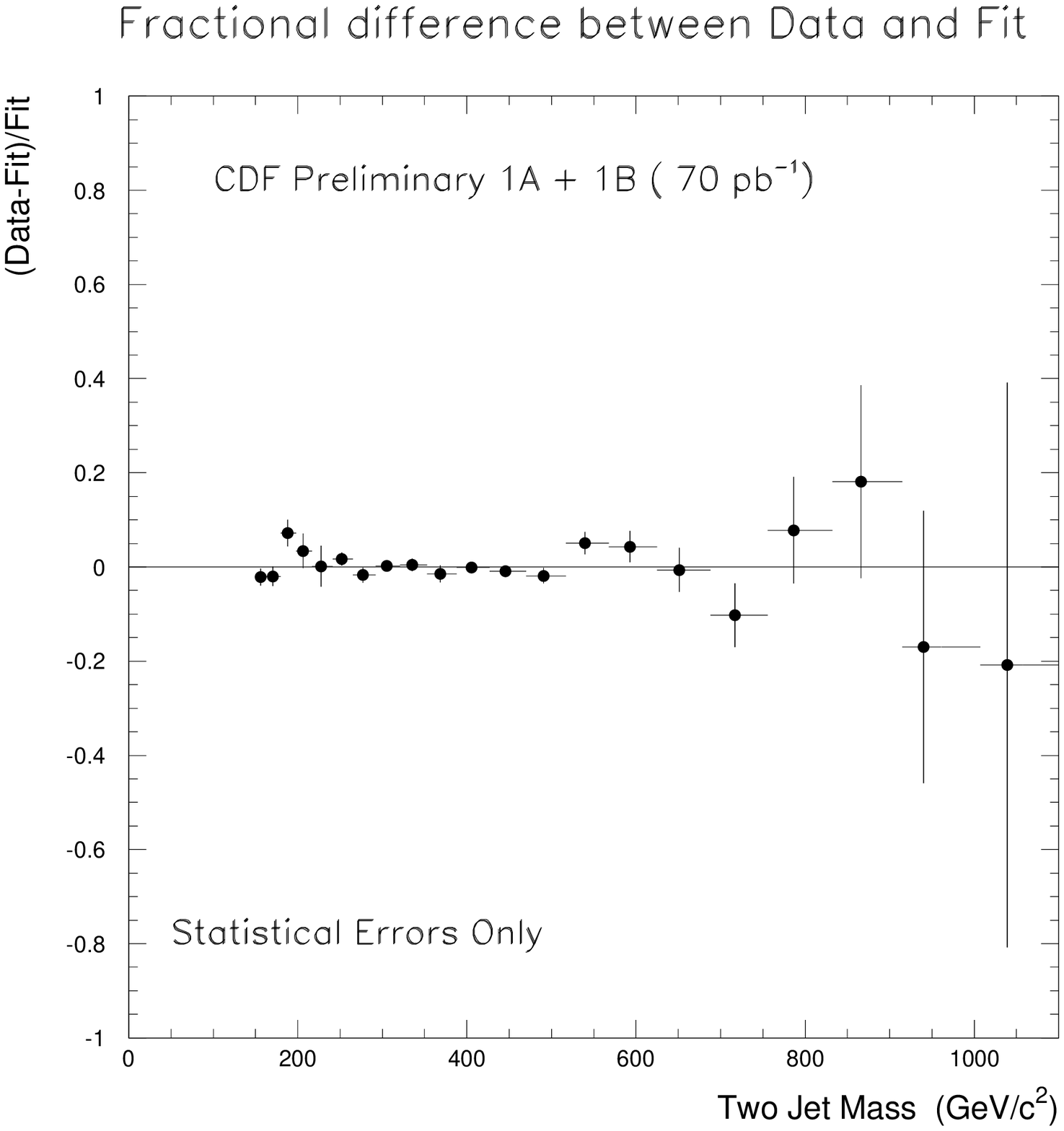}
\caption{ \large The dijet mass data (solid points) is compared to
a parameterization fit to the data (curve). The logarithmic plot also
shows a QCD simulation (open boxes).}
\label{fig_dijet}
\end{figure}
\noindent
.089 pb$^{-1}$, 1.92 pb$^{-1}$, 9.52 pb$^{-1}$,
and 69.8 pb$^{-1}$ respectively.  Jets are defined with a fixed cone clustering
algorithm
(R=0.7) and then corrected for detector response, energy lost outside
the cone, and underlying event.  We take the two highest $P_T$ jets and
require that they have pseudorapidity $|\eta|<2$ and a CMS scattering angle
$|\cos\theta^*| = |\tanh[(\eta_1-\eta_2)/2]| < 2/3$. The $\cos\theta^*$ cut
provides
uniform acceptance as a function of mass and reduces the QCD background which
peaks at $|\cos\theta^*|=1$.
In Fig.~\ref{fig_dijet} the dijet mass distribution is presented
as a differential cross section in bins of the mass resolution
($\sigma\sim 10$\%). At high mass the data is systematically higher than a
prediction from PYTHIA plus a CDF detector simulation, similar to the inclusive
jet $E_T$ spectrum~\cite{ref_anwar}. To search for new particles we determine
the
QCD background by fitting the data to a smooth function of
three parameters~\cite{ref_param}; Fig.~\ref{fig_dijet} shows the fractional
difference between the data and the fit ($\chi^2/DF=1.43$). We note
upward fluctuations near 200 GeV/c$^2$ ($2.4\sigma$), 550 GeV/c$^2$
($2.6\sigma$) and
850 GeV/c$^2$ ($1\sigma$).

\begin{figure}[tbh]
\epsfysize=3.3in
\epsffile[36 200 400 590]{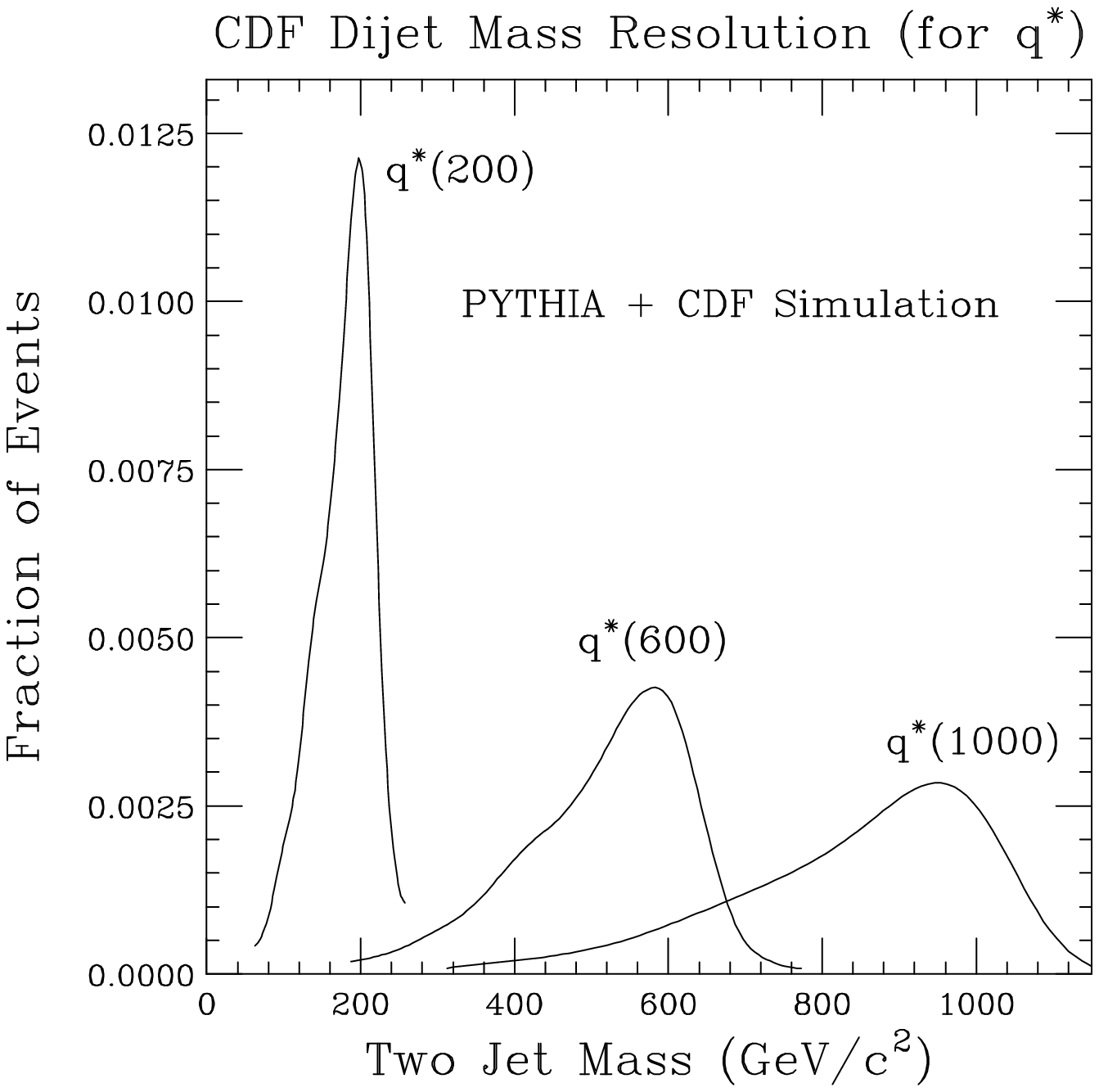}
\vspace*{-3.3in}
\hspace*{3.4in}
\epsfysize=3.3in
\epsffile[30 144 540 680]{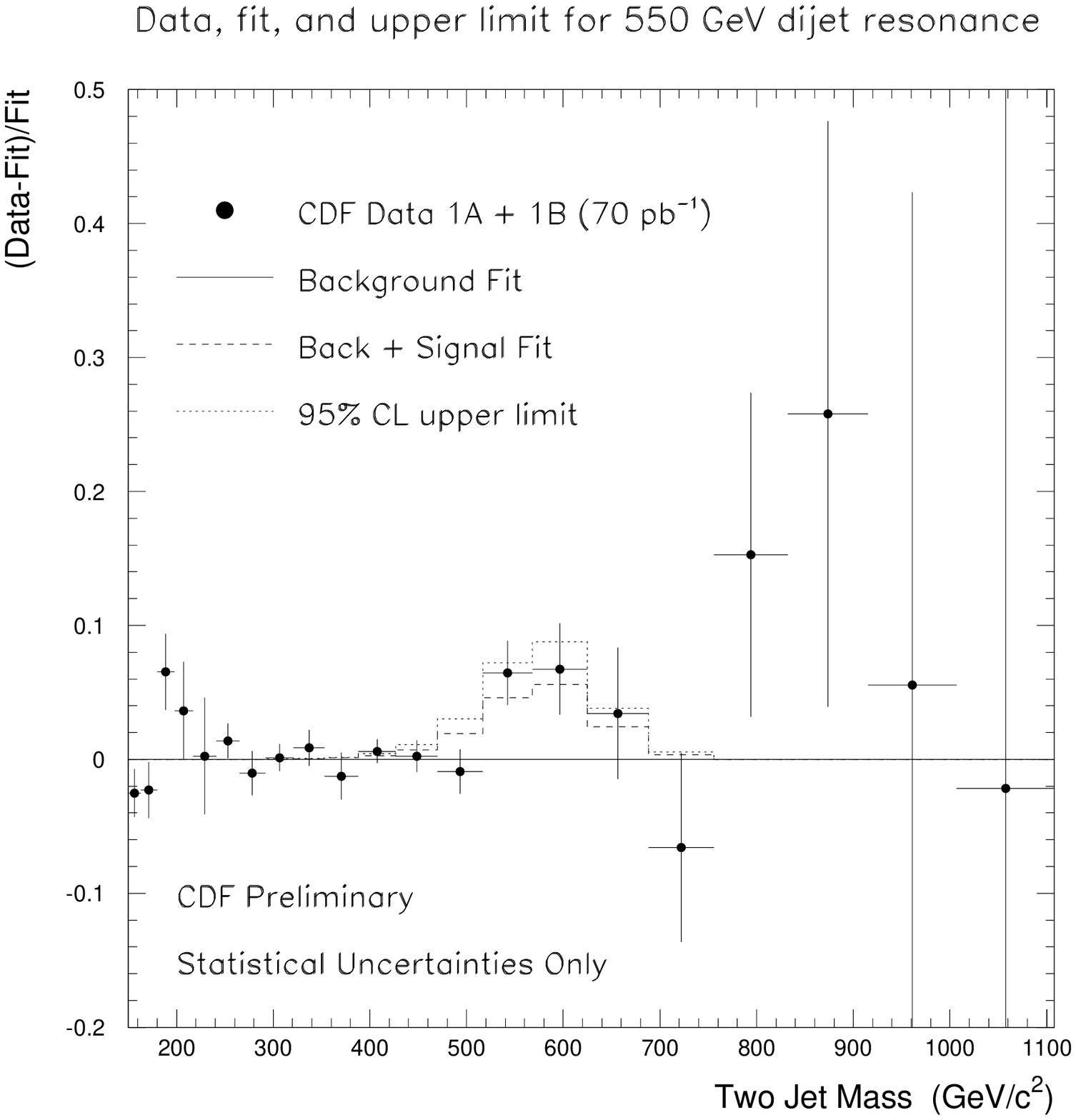}
\caption{ \large left: CDF dijet mass resolution for narrow resonances like
excited
quarks, including the effects of radiation
and detector resolution.
right: The dijet mass data (solid points) fit with a background (solid line)
and a 550 GeV/c$^2$ resonance (dashed hist).}
\label{fig_resonance}
\end{figure}

For narrow resonances it is sufficient to determine the mass
resolution for only one type
of new particle because the detector resolution dominates the width.
In Fig.~\ref{fig_resonance} we show the mass resolution
for excited quarks (q*) from PYTHIA plus a CDF detector simulation; the long
tail at low mass comes from gluon radiation.
For each value of new particle mass in 50 GeV/c$^2$ steps, we perform a binned
maximum likelihood fit of the data to the background
parameterization and the mass
resonance shape. In Fig.~\ref{fig_resonance} we display the best fit and
95\% confidence level upper limit for a 550 GeV/c$^2$ resonance. For the mass
region $517<M<625$ GeV/c$^2$, there are 2947 events in the data, $2810\pm 53$
events ($2.6 \sigma$) in the background for the fit without a resonance,
$2765\pm53$ events ($3.4 \sigma$) in the background for the fit that includes
the resonance, and the value of the resonance cross section from the fit is
$5.8\pm 2.9$ pb (statistical).

\begin{figure}[b]
\vspace*{-2in}
\epsfysize=3.3in
\epsffile[36 144 540 680]{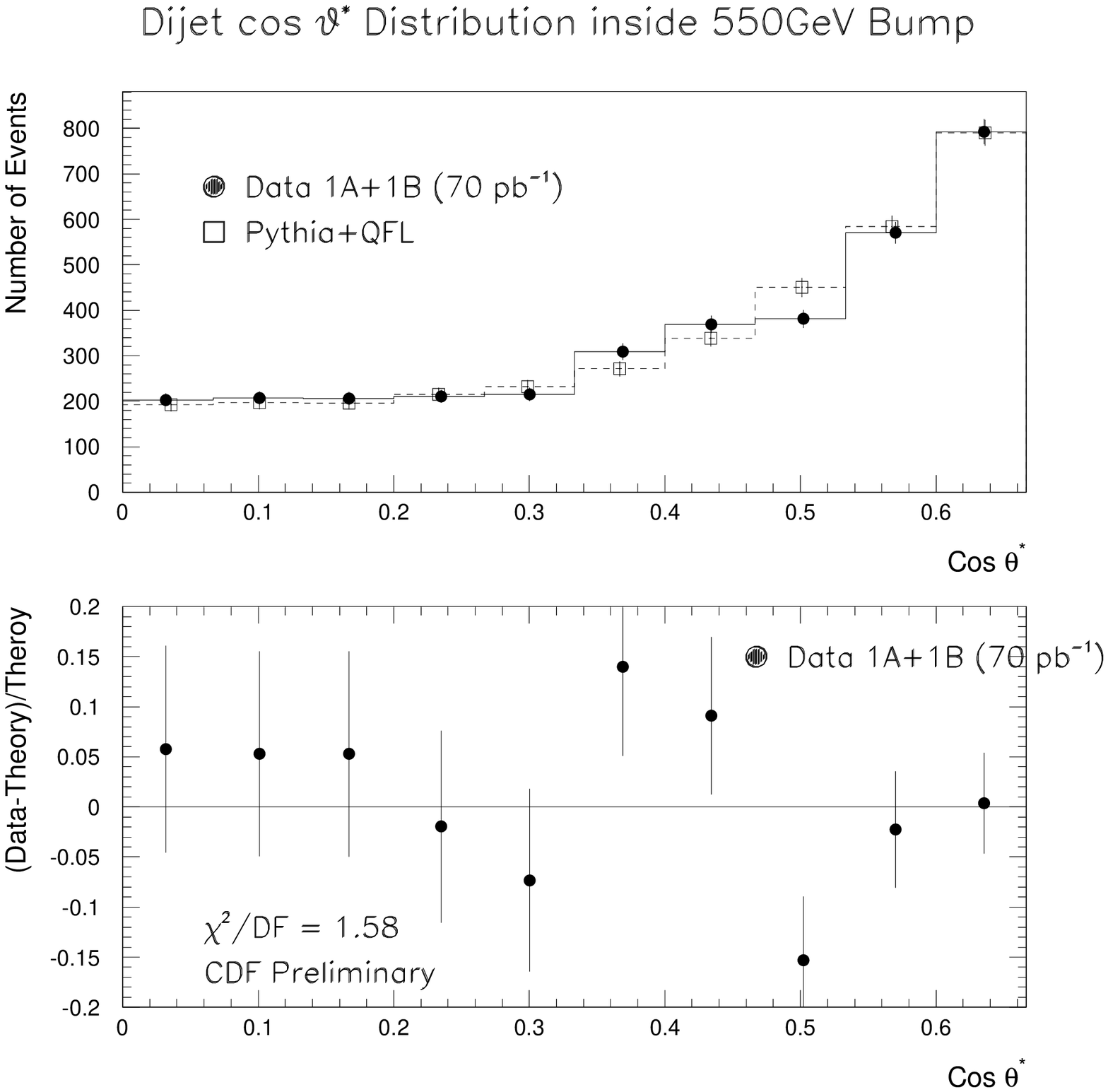}
\vspace*{-3.3in}
\hspace*{3.4in}
\epsfysize=3.3in
\epsffile[36 144 540 680]{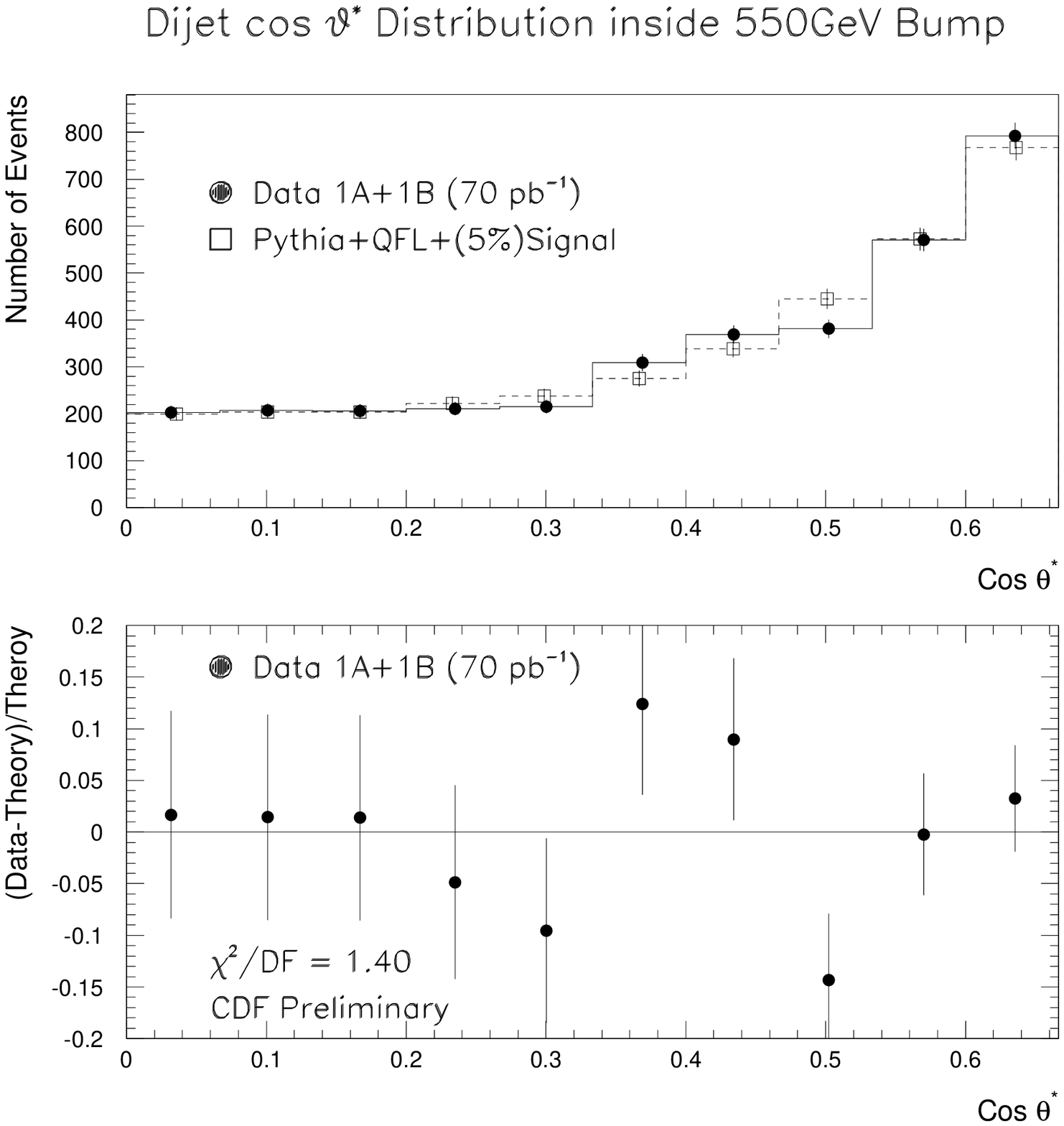}
\caption{ \large left: The angular distribution of data in the two central bins
of
the 550 GeV/c$^2$ fluctuation(solid points) is compared to a QCD simulation
(open boxes) and the fractional difference is shown beneath. right: same as
left but fit to a QCD simulation plus a floating signal. The predictions
are normalized to the data.}
\label{fig_cos}
\end{figure}

In Fig.~\ref{fig_cos} we study the angular distribution of the fluctuation
in the mass region $517<M<625$ GeV/c$^2$.  The angular distribution is
compatible
with both QCD alone, and with
\clearpage
\noindent QCD + 5\% excited quark (best fit).  This
amount of excited quark is coincidentally the same as found in the mass fit.
Although the fluctuation is interesting, we conclude it is not yet
statistically significant, and proceed to set limits on new particle
production.

\clearpage

\begin{figure}[tbh]
\epsfysize=3.3in
\epsffile[30 144 540 680]{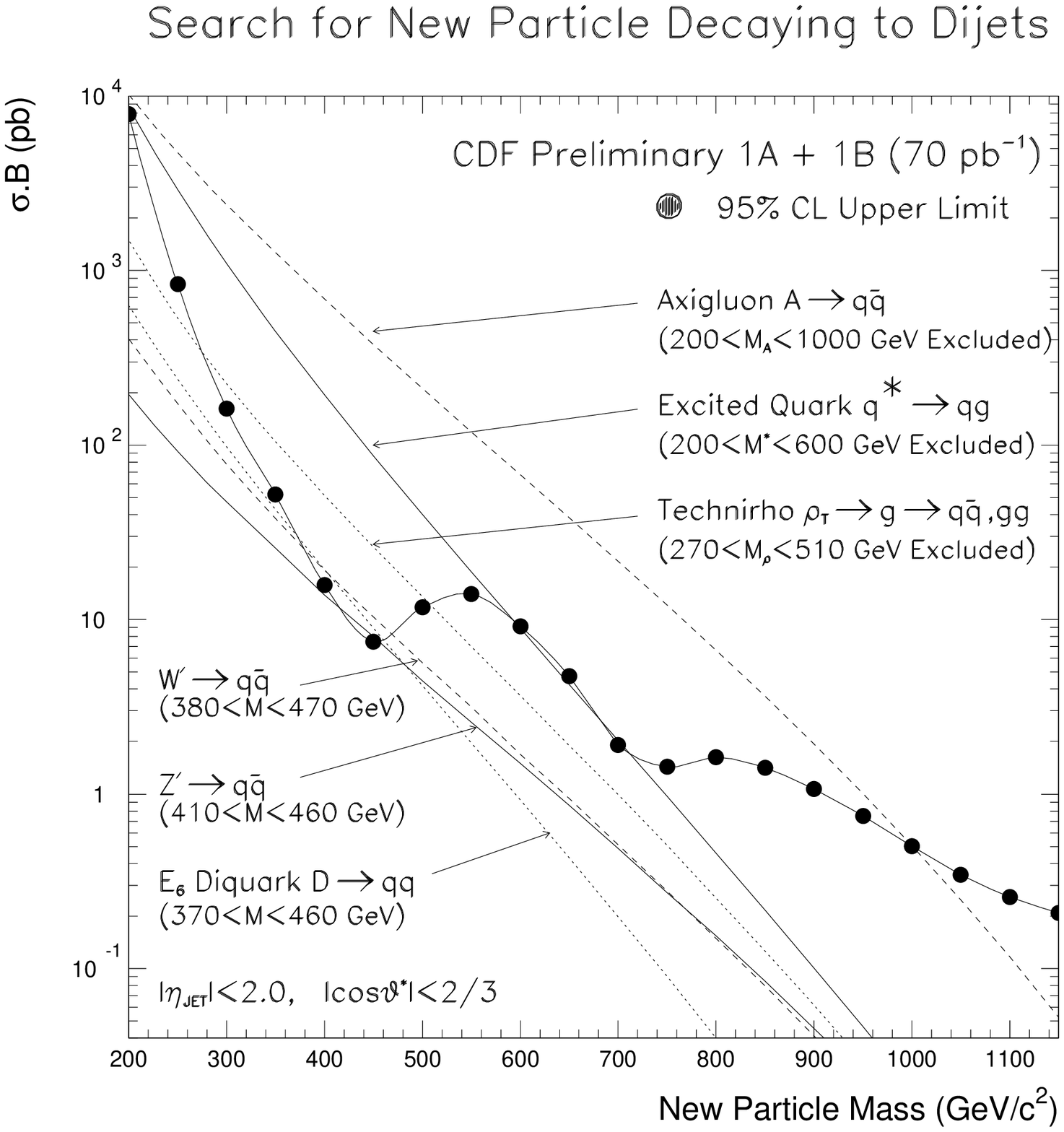}
\vspace*{-3.3in}
\hspace*{3.4in}
\epsfysize=3.3in
\epsffile[36 190 540 720]{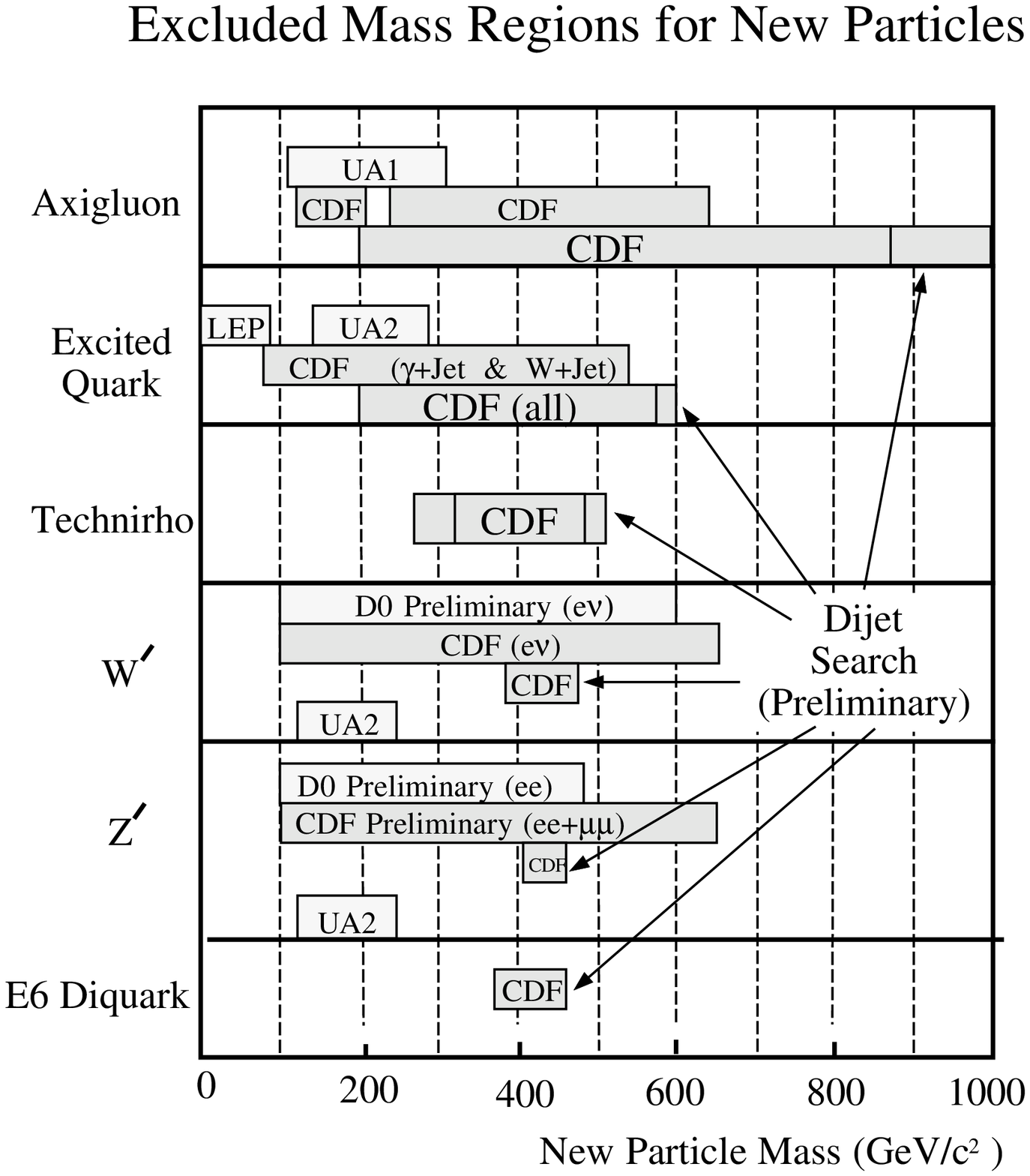}
\caption{ \large left: Upper limits on the cross section for new particles.
right: Excluded mass regions compared to previous searches.}
\label{fig_dijet_limit}
\end{figure}

\vspace*{-0.05in}
{}From the likelihood distribution including experimental systematic
uncertainties~\cite{ref_dijet} we obtain the 95\% CL upper limit on the
cross section for new particles shown in Fig.~\ref{fig_dijet_limit}.
We compare this to the cross
section for axigluons (excluding $200<M<1000$ GeV/c$^2$), excited
quarks (excluding $200<M<600$ GeV/c$^2$), technirhos (excluding $270<M<510$
GeV/c$^2$),
W$^\prime$ (excluding $380<M<470$ GeV/c$^2$), Z$^\prime$ (excluding
$410<M<460$ GeV/c$^2$),
and E6 diquarks (excluding $370<M<460$ GeV/c$^2$). The calculations are lowest
order~\cite{ref_gauge} using CTEQ2L parton distributions~\cite{ref_CTEQ} and
one-loop $\alpha_s(m^2)$ and require $|\eta|<2$ and $|\cos\theta^*|<2/3$.

\section{\large \bf Topcolor}

The large mass of the top quark suggests that the third generation may
be special.  Topcolor~\cite{ref_topc1,ref_topc2} assumes that the
top mass is large mainly because of a dynamical $t\bar{t}$ condensate generated
by a new strong dynamics coupling to the third generation. Here the $SU(3)_C$
of QCD is a low energy symmetry arising from the breaking of an $SU(3)_1$
coupling to the
third generation and an $SU(3)_2$ coupling to the first two generations only.
There are then massive color octet bosons, topgluons $B$, which couple
largely to $b\bar{b}$ and $t\bar{t}$.  The topgluon is strongly
produced and decays mainly to the third generation
($q\bar{q} \rightarrow B \rightarrow b\bar{b}, t\bar{t}$) with a relatively
large natural width ($\Gamma\ge 0.11M$).  Here we search for the topgluon
in the $b\bar{b}$ channel.

An additional $U(1)$ symmetry is
introduced~\cite{ref_topc2} to keep the $b$ quark light while the top quark is
heavy; this leads to a topcolor $Z^{\prime}$, which again
couples largely to $b\bar{b}$ and $t\bar{t}$.
The topcolor $Z^{\prime}$ is electroweakly produced and decays mainly to the
third generation ($q\bar{q} \rightarrow  Z^{\prime} \rightarrow b\bar{b},
t\bar{t}$) with a narrow natural width ($\Gamma\ge 0.012M$).  Here we search
for the topcolor $Z^{\prime}$ in both the the $b\bar{b}$ and $t\bar{t}$
channel; the $t\bar{t}$ channel is the most sensitive because the coupling to
$t\bar{t}$ is larger.
\clearpage
\vspace*{-0.2in}
\section{\large \bf Search for New Particles Decaying to $b\bar{b}$}
\begin{figure}[h]
\epsfysize=2.6in
\epsffile[36 170 580 615]{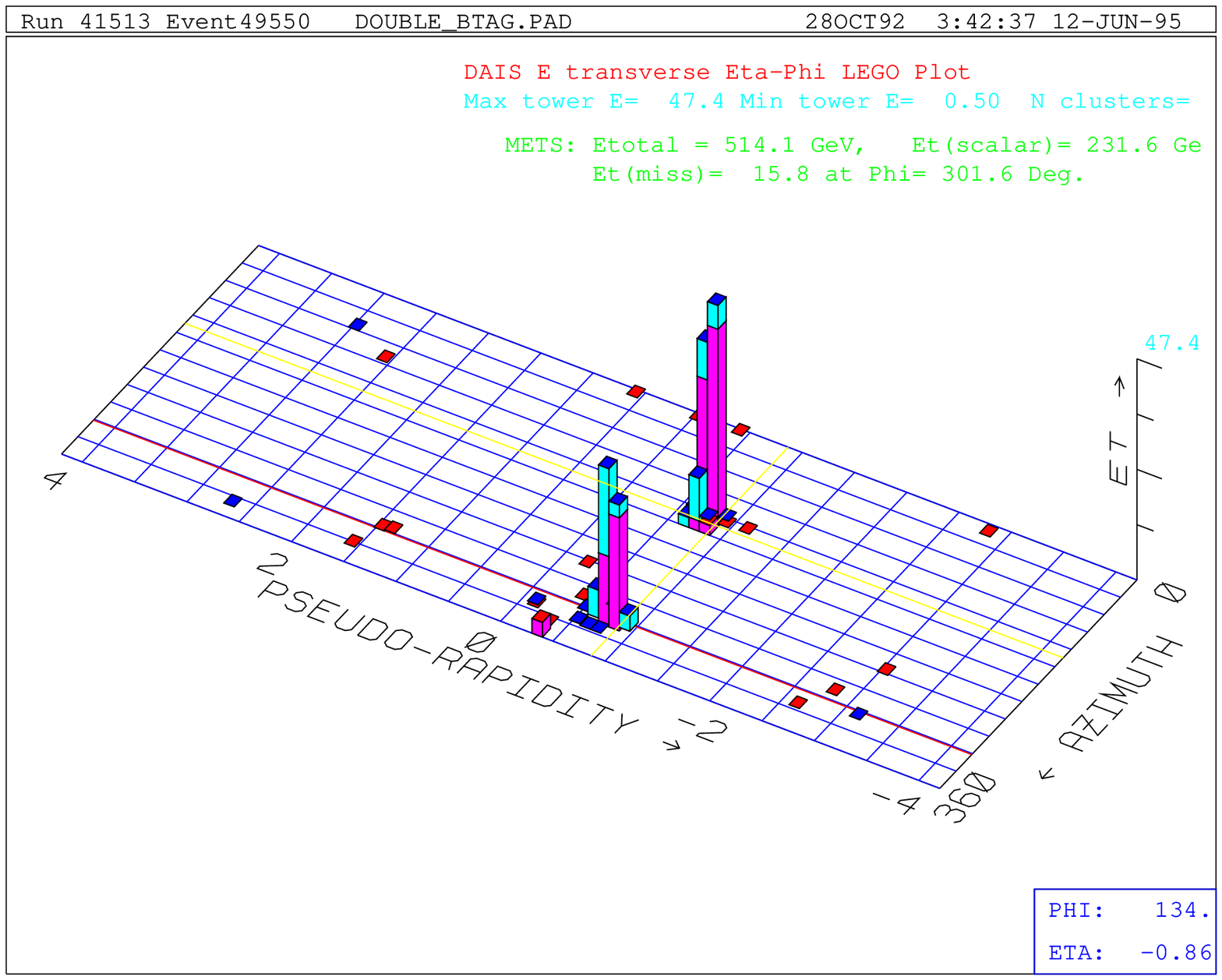}
\vspace*{-2.6in}
\hspace*{3.3in}
\epsfysize=2.6in
\epsffile[36 170 580 615]{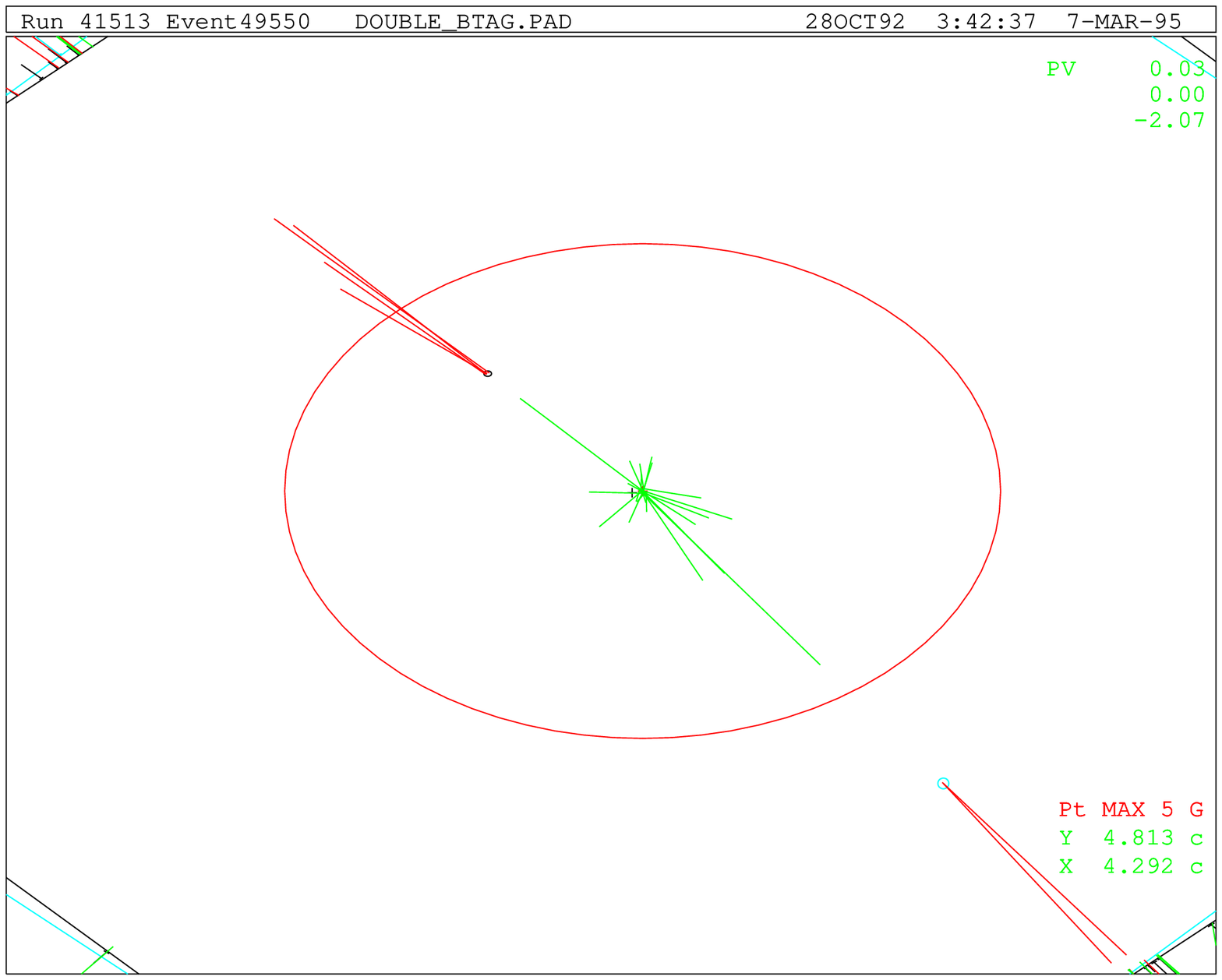}
\caption{ \large The highest mass (256 GeV/c$^2$) double b-tagged dijet
event. left: Lego plot showing the two jets. right: Vertex plot
showing the SVX tracks forming displaced secondary vertices: one inside the
beam pipe, the other outside.}
\label{fig_btag_event}
\end{figure}

We start with the dijet search in 19 pb$^{-1}$ of run 1A data~\cite{ref_dijet}
and additionally require at least one of the two leading jets be tagged
as a bottom quark.  The b-tag requires a displaced vertex in the the
secondary vertex detector~\cite{ref_top_evidence}. The $b\bar{b}$ event
efficiency is $25\pm 2$\% independent of dijet mass. From fits to the
$c\tau$ distribution, we estimate that the sample is roughly 50\% bottom,
30\% charm, and 20\% mistags of plain jets.  PYTHIA predicts that 1/5 of these
bottom quarks are direct $b\bar{b}$, and the rest are from gluon splitting
and flavor excitation.  Consequentially, only about $1/5 \times 50$\% $=$ 10\%
of our sample is direct $b\bar{b}$. We expect both the purity and
efficiency to increase when we use the run 1B dataset and a new tagging
algorithm~\cite{ref_top_discovery}.  With higher tagging efficiency we should
be
able to make better use of double b-tagged events like the one in
Fig.~\ref{fig_btag_event}.
\vspace*{-0.05in}

\begin{figure}[t]
\vspace*{-0.05in}
\epsfysize=3.2in
\epsffile[30 144 540 680]{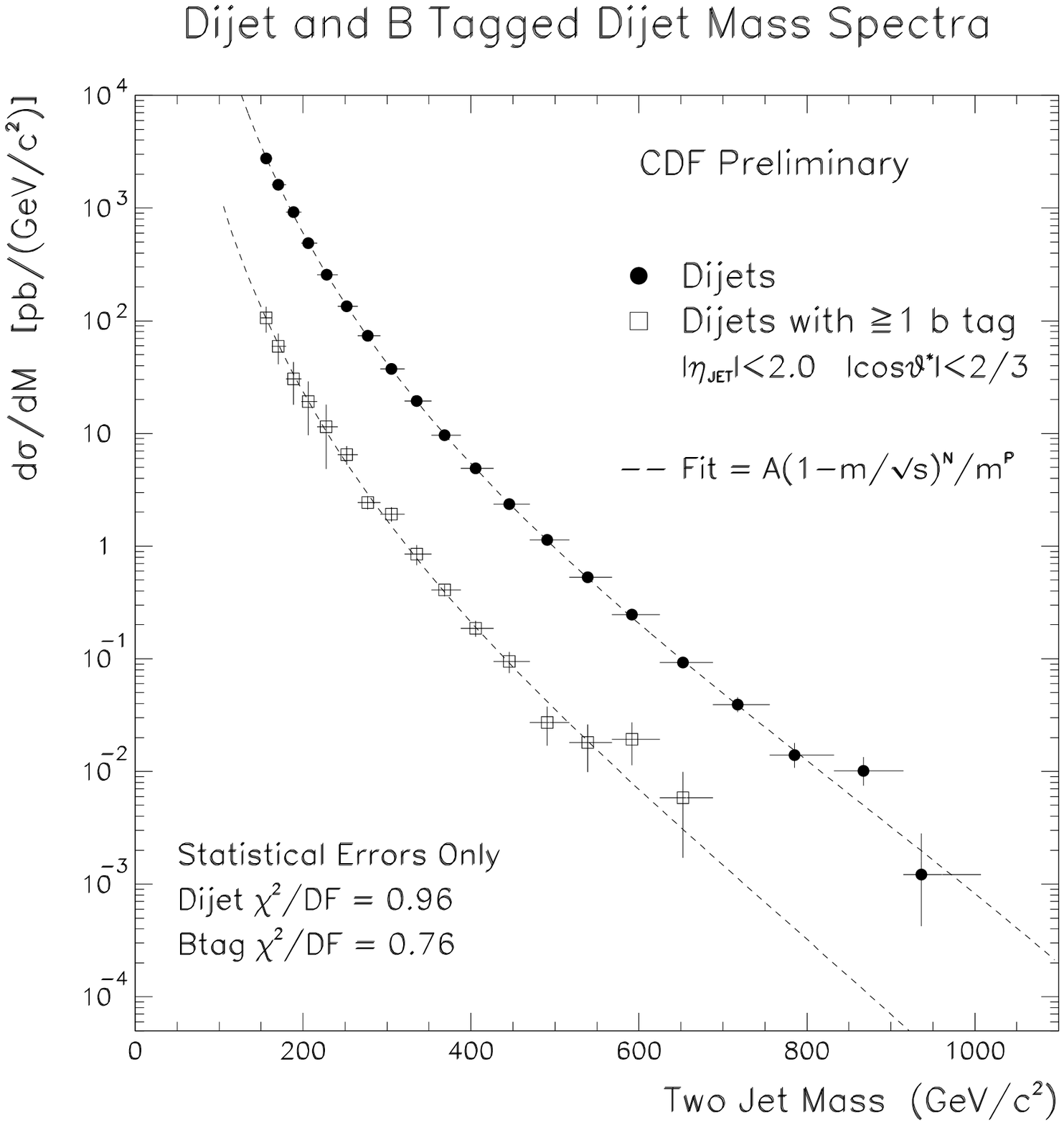}
\vspace*{-3.2in}
\hspace*{3.3in}
\epsfysize=3.2in
\epsffile[36 144 540 680]{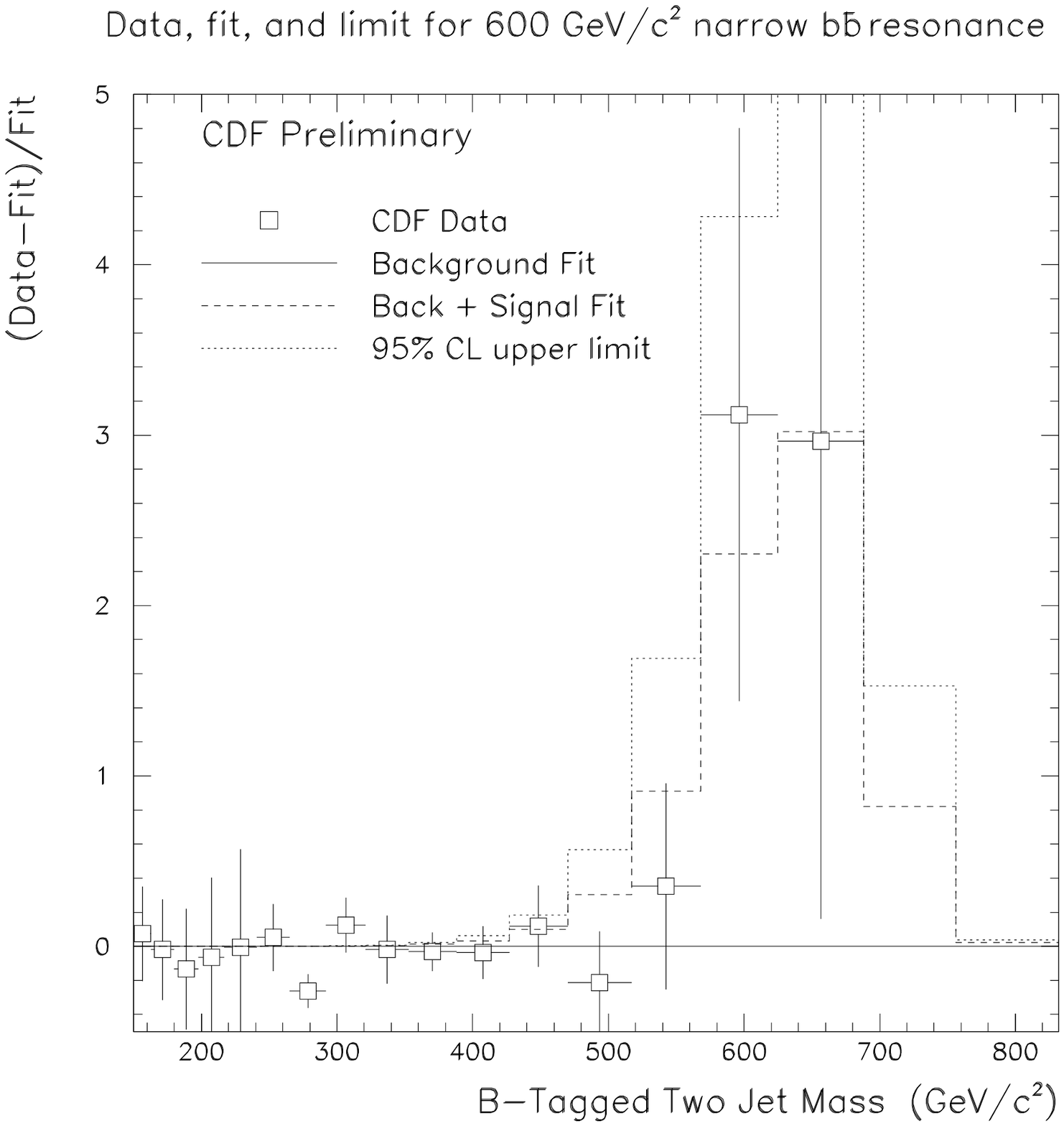}
\caption{ \large left: Dijet mass data (solid points) and b-tagged dijet mass
data (open boxes) both from run 1A only, are compared to a parameterization fit
to the data (curve).
right: The b-tagged dijet mass data (open boxes) fit by a background (solid
line) and a 600 GeV/c$^2$ narrow resonance (dashed hist).}
\label{fig_btag}
\end{figure}

In Fig.~\ref{fig_btag} we show the b-tagged dijet mass distribution corrected
for the $b\bar{b}$ efficiency.  Also shown is the untagged dijet mass
distribution from run 1A, and both are well fit with our standard
parameterization~\cite{ref_param}.  The b-tagged dijet data has an
upward fluctuation near 600 GeV/c$^2$.  We model the shape of a narrow
resonance using PYTHIA Z$^{\prime}$ production and a CDF detector simulation.
In Fig.~\ref{fig_btag} we fit the b-tagged data to a 600 GeV/c$^2$ narrow
resonance,
and find a cross section of $3.2\pm 1.6$ pb (statistical).  Note that this
is comparable to the dijet fluctuation in both mass and rate. However, there
are only 8 events in the last two data bins of Fig.~\ref{fig_btag}, and the
fluctuation is only a $2\sigma$ effect, so we proceed to set limits on new
particle production.

\begin{figure}[b]
\vspace*{-2in}
\epsfysize=3.0in
\epsffile[36 200 400 600]{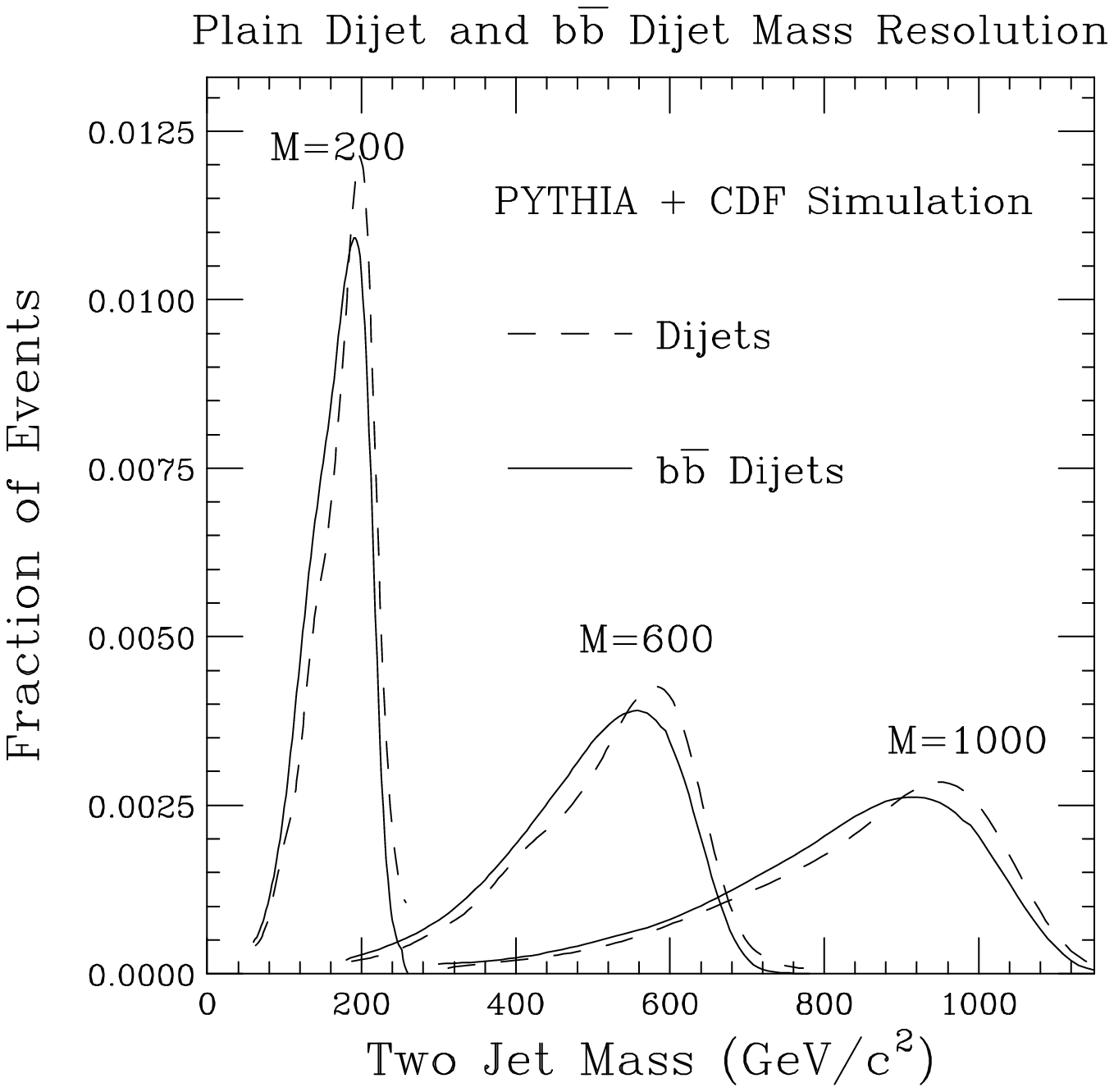}
\vspace*{-3.0in}
\hspace*{3.25in}
\epsfysize=3.0in
\epsffile[36 200 400 600]{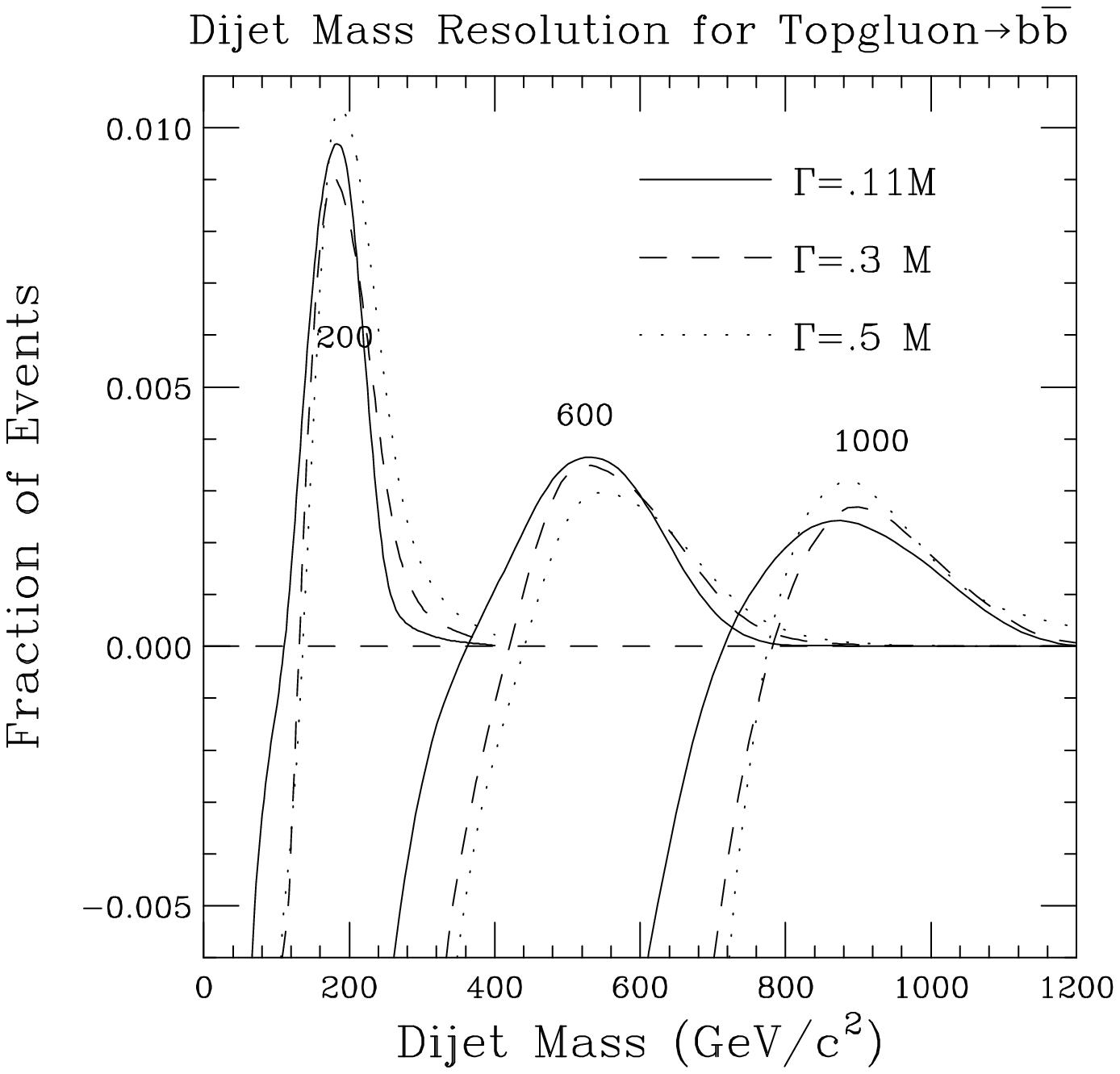}
\caption{ \large left: $b\bar{b}$ mass resolution for narrow resonances like
topcolor
Z$^{\prime}$, including the effects of radiation
and detector resolution.
right: $b\bar{b}$ mass resolution for topgluons, including interference with
$g\rightarrow b\bar{b}$, are shown for three different topgluon widths.}
\label{fig_b_bbar_res}
\end{figure}

\begin{figure}[t]
\epsfysize=7.7in
\epsffile[40 100 570 760]{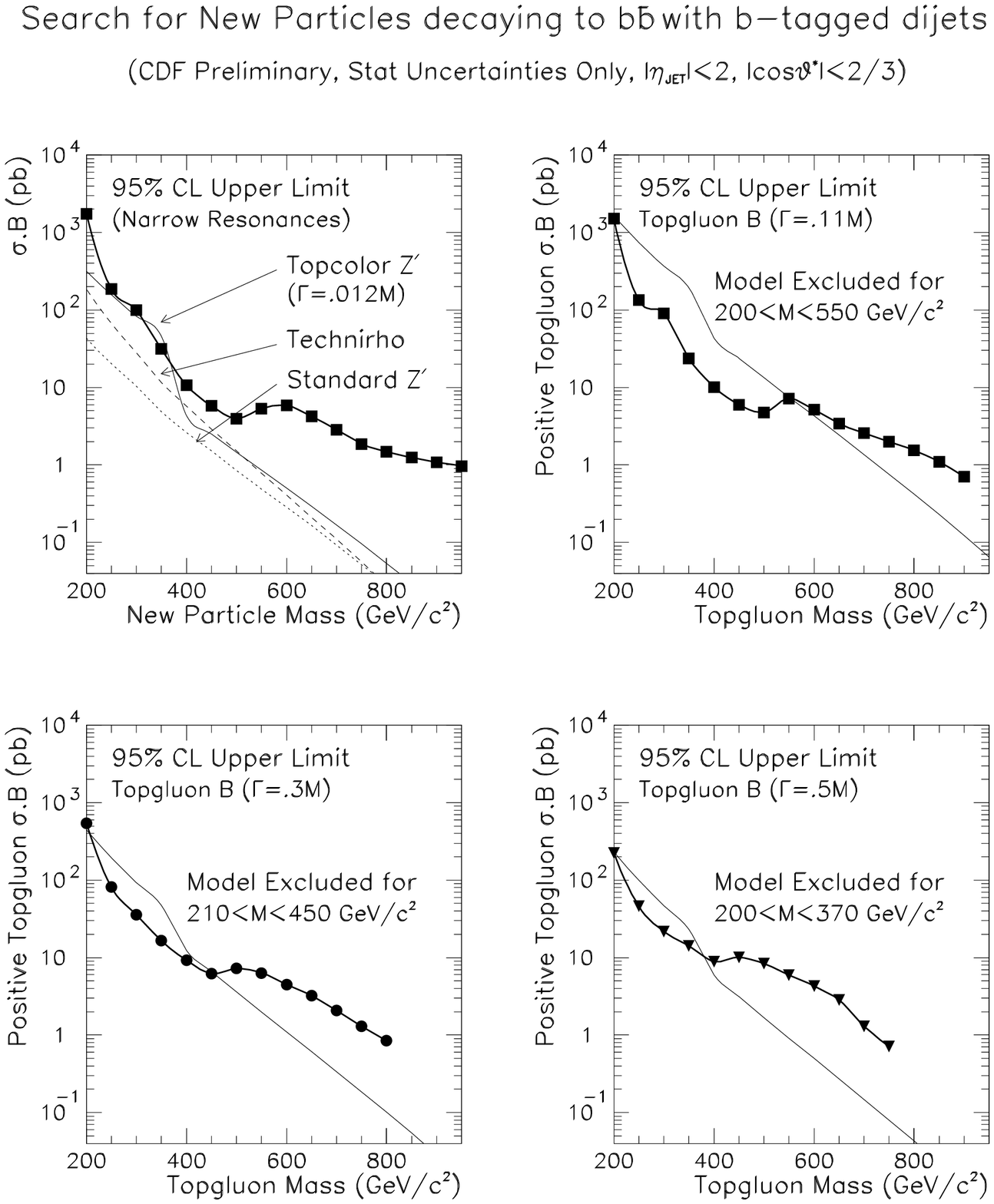}
\caption{ \large upper left: Preliminary upper limits on the cross section for
a
narrow resonances decaying to $b\bar{b}$ (squares) are compared to the
theoretical prediction for topcolor Z$^{\prime}$, color octet technirhos,
and a standard Z$^{\prime}$. remaining plots: Upper limits on cross section for
topcolor bosons of fractional width $0.11$ (squares), $0.3$ (circles) and
$0.5$ (triangles) are compared with theoretical predictions for topgluons;
excluded mass regions are listed.  What is plotted is the positive topgluon
cross section and limits: the cross section in that region where the topgluon
produces an excess above QCD (there is a region where the destructive
interference between the topgluon and the normal gluon drives the total cross
section beneath the standard QCD prediction; see fig.~\ref{fig_b_bbar_res}).}
\label{fig_btag_limit}
\end{figure}

	We perform two kinds of fits for the limits.  First, narrow resonances
are modelled as described above, and the mass resolution in the CDF detector
is shown in Fig.~\ref{fig_b_bbar_res}.
Second, wide resonances characteristic of
topgluons~\cite{ref_topgluon_shape},
including interference with normal gluons, was incorporated into PYTHIA and a
CDF detector simulation.  The mass resolution in Fig.~\ref{fig_b_bbar_res}
displays destructive interference to the left of the resonance; models with
destructive interference on the right side of the resonance will be considered
in the future.
\clearpage
Limits on new particle production are shown in
Fig.~\ref{fig_btag_limit}.
The theoretical cross sections are lowest order and use
CTEQ2L parton distributions.
For narrow resonances the production cross sections aren't large enough for
us to set mass limits at this time.  For topgluons the production cross
sections~\cite{ref_private}  are larger, and we are able to exclude
at 95\% CL topgluons of width
$\Gamma=0.11M$ in the mass region $200<M<550$ GeV/c$^2$,
$\Gamma=0.3M$ for $210<M<450$ GeV/c$^2$, and
$\Gamma=0.5M$ for $200<M<370$ GeV/c$^2$.
\clearpage

\clearpage

\section{\large \bf Search for New Particles Decaying to $t\bar{t}$}

To search for new particles decaying to $t\bar{t}$ we start with the data
sample from the top mass measurement~\cite{ref_top_discovery}.
There we used top decays to W + four jets with at least one b-tag, and found
19 events on a background of $\sim 7\pm 2$, resulting in a top mass of
$176\pm 8$(stat)$\pm 10$(sys) GeV/c$^2$.  That analysis fit the entire event
for the top hypothesis, discarding events with $\chi^2>10$ (poor fit).  Here
we add the additional constraint that the top mass is 176 GeV/c$^2$, which
significantly enhances our resolution of the $t\bar{t}$ mass.  Two of the 19
events fail the $\chi^2>10$ cut when the top mass constraint is added to the
fit, leaving us with 17 events.

\begin{figure}[tbh]
\epsfysize=3.1in
\epsffile[18 144 520 650]{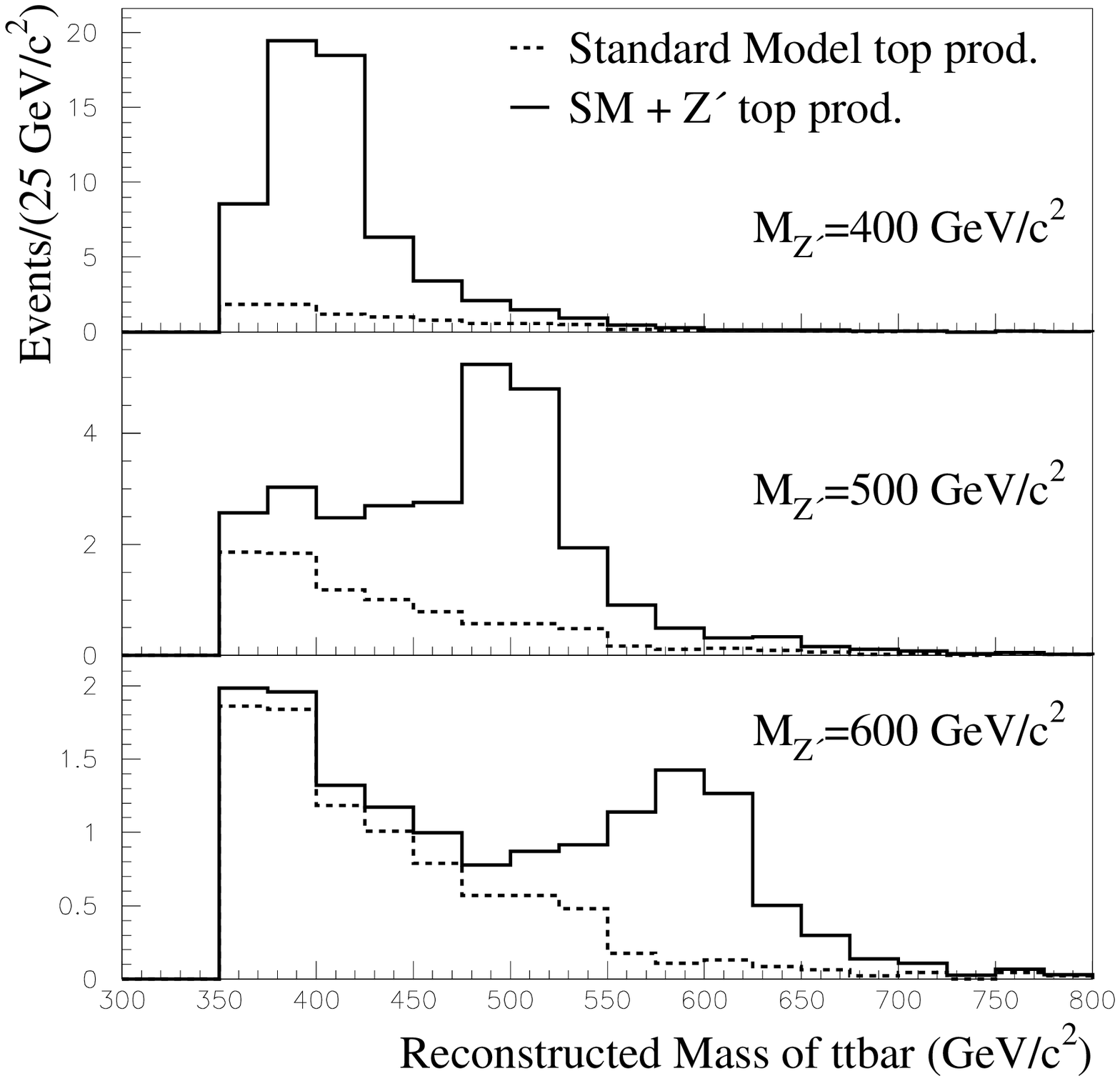}
\vspace*{-3.1in}
\hspace*{3.2in}
\epsfysize=3.1in
\epsffile[18 144 520 650]{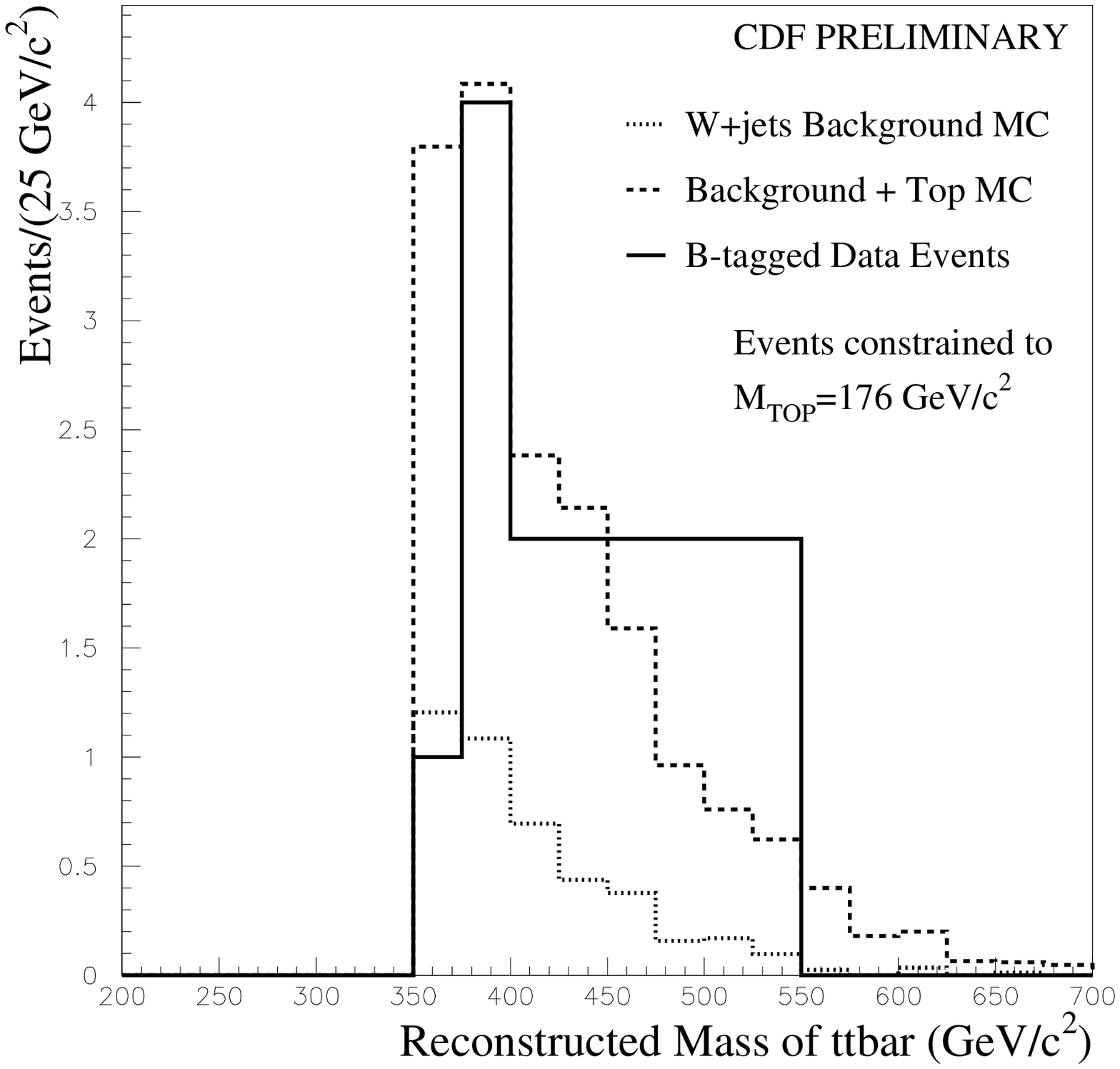}
\caption{ \large left: Simulation of SM $t\bar{t}$ candidates (dashes) compared
to
SM + topcolor Z$^{\prime}$ (solid) for 3 different Z$^{\prime}$ masses.
right: $t\bar{t}$ candidate data (solid) compared to W+jets background MC
(dotted) and total standard model background including $t\bar{t}$ (dashed).
The simulations and data are for 67 pb$^{-1}$.}
\label{fig_ttbar}
\end{figure}

The $t\bar{t}$ mass distribution expected from a narrow resonance, normalized
to the topcolor $Z^{\prime}$ predicted rate~\cite{ref_private}, is shown in
Fig.~\ref{fig_ttbar}.  Here we used PYTHIA $Z^{\prime}\rightarrow t\bar{t}$.
Also in Fig.~\ref{fig_ttbar} is the Monte Carlo
distribution of the background, on the left standard model top production
from HERWIG, and on the right QCD W + jets background from VECBOS with parton
showers from HERWIG. All Monte Carlos include a CDF detector simulation.
On the left in Fig.~\ref{fig_ttbar},
the comparison of the topcolor Z$^{\prime}$ to SM $t\bar{t}$ simulations
illustrates that in this data sample we are sensitive to topcolor
$Z^{\prime}$ up to a mass of roughly 600 GeV/c$^2$.
Finally, on the right in
Fig.~\ref{fig_ttbar}, we present the $t\bar{t}$ candidate mass distribution
from CDF compared to the total standard model prediction. Given the
statistics the agreement is quite good overall. The small shoulder of
6 events on a background of $2.4$ in the region $475<M<550$ GeV/c$^2$ is
in an interesting mass region, given the dijet and $b\bar{b}$ search results,
but is not statistically significant.  Upper limits on the $t\bar{t}$ cross
section as a function of $t\bar{t}$ mass, and on a topcolor $Z^{\prime}$, are
currently in progress.

\section{\large \bf Conclusions}
We have searched for new particles decaying to dijets, $b\bar{b}$, and
$t\bar{t}$.  In the dijet channel we set the most significant direct mass
exclusions to date on the hadronic decays of axigluons
(excluding $M<1000$ GeV/c$^2$), excited quarks (excluding $M<600$ GeV/c$^2$),
technirhos (excluding $270<M<510$ GeV/c$^2$),
W$^{\prime}$ (excluding $380<M<470$ GeV/c$^2$),
Z$^\prime$ (excluding $410<M<460$ GeV/c$^2$),
and for the first time E6 diquarks (excluding $370<M<460$ GeV/c$^2$).  In the
$b\bar{b}$ channel we set the first limits on topcolor, excluding a model
of topgluons for width
$\Gamma=0.11M$ in the mass region $200<M<550$ GeV/c$^2$,
$\Gamma=0.3M$ for $210<M<450$ GeV/c$^2$, and
$\Gamma=0.5M$ for $200<M<370$ GeV/c$^2$.
The search for topcolor in the $t\bar{t}$ channel has just begun and limits
are in progress.

Limits are only a consolation prize; the main emphasis of our search is to
explore the possibility of a signal.
Although we do not have significant evidence for new particle
production, the $500-600$ GeV/c$^2$ region shows upward fluctuations in all
three channels.
We cannot ignore the exciting possibility that these apparently
separate fluctuations may be the first signs of a new physics beyond the
standard model. The remaining integrated luminosity for run 1B, currently
being accumulated and analyzed, has the potential to either kill the
fluctuations or reveal what may be the most interesting new physics in
a generation.\\

%



\begin{references}
\bibitem{ref_gauge} For new gauge bosons we use a K-factor to account
for higher order terms.
\end{references}
\end{document}